\documentclass[runningheads]{llncs}
\usepackage{cite}
\usepackage{amsmath,amssymb,amsfonts}
\usepackage{algorithmic}
\usepackage{graphicx}
\usepackage{textcomp}
\usepackage{xcolor}
\usepackage{bmpsize}
\usepackage{float}
\usepackage{lipsum}
\usepackage{appendix}

\usepackage[colorlinks=true,urlcolor=black]{hyperref}
\usepackage[ruled,vlined,linesnumbered]{algorithm2e}

\makeatletter
\newcommand\notsotiny{\@setfontsize\notsotiny\@vipt\@viipt}
\makeatother
\begin{document}
\title{Process Mining Algorithm for Online Intrusion Detection System}
%
%
\author{Yinzheng Zhong\orcidID{0000-0001-8477-3956} \and
John Y. Goulermas\orcidID{0000-0003-0381-124X} \and
Alexei Lisitsa\orcidID{0000-0002-3820-643X}}
%
%
\institute{Department of Computer Science, University of Liverpool, UK \email{sgyzhon5@liverpool.ac.uk,
goulerma@liverpool.ac.uk,
a.lisitsa@liverpool.ac.uk}\\
}
\maketitle              
\begin{abstract}
In this paper, we consider the applications of process mining in intrusion detection. We propose a novel process mining inspired algorithm to be used to preprocess data in intrusion detection systems (IDS). The algorithm is designed to process the network packet data and it works well in online mode for online intrusion detection. To test our algorithm, we used the CSE-CIC-IDS2018 dataset which contains several common attacks. The packet data was preprocessed with this algorithm and then fed into the detectors. We report on the experiments using the algorithm with different machine learning (ML) models as classifiers to verify that our algorithm works as expected; we tested the performance on anomaly detection methods as well and reported on the existing preprocessing tool CICFlowMeter for the comparison of performance.

\end{abstract}
\section{Introduction}
With the growth of applications that relies on internet communication, cybercrime became a serious issue that affects many areas. By estimation, about 33 billion records of personal information including addresses, credit card information, or social security numbers etc., will be stolen in 2033 \cite{symantec}.
The intrusion detection systems (IDS) protect computer systems by monitoring the network or system activities. One of the main challenges in the design of IDS is to have fast and robust methods for network traffic assessment to be used for the detection of attacks and malicious behaviour.
The \emph{process mining} has risen recently as a promising research direction aiming at systematic developments of the methods for building behavioural or workflow models from event logs \cite{van2004process, van2011process}. The process mining is essentially approaches that takes information (e.g. cases, timestamps and events) from the event logs for building the workflow models (process models) which can then be used for analytical tasks. The process model describes the transitions of events within traces.

While the applications of process mining in the security have been considered, e.g. in \cite{van2005process}, its applications in IDS remain largely unexplored. In this paper, we propose process mining inspired technique to be used at the preprocessing stage to generate a behaviour model, which subsequently be classified as attack/no attack or normal/malicious behaviour by trained machine learning models.
There are similar approaches for IDS, for example, based on \emph{data mining} \cite{lee1998data} and \emph{machine learning} \cite{agarap2018neural, kim2019intrusion}. In most of the cases, however, these approaches can only detect the threats after features been generated based on flows. In our approach, the process mining is used as the preprocessing step, while machine learning is used as the classifier. 
Our proposed algorithm for process mining of network data can be seen as a modification of 
the initial model mentioned in the fuzzy mining algorithm \cite{gunther2007fuzzy}. The latter was modified for better online processing, and techniques such as aggregation and abstraction in fuzzy mining could also be applied. The rest of the paper is organized as follows. In the next section, we give a short outline of the related work. After that, in Section \ref{section:frequency} we present the proposed process mining inspired algorithm. The setup for machine learning is discussed in Section \ref{section:ml}. Section \ref{section:results} reports on experiments and Section \ref{section:conclusion} presents the discussion and outlines the future work. 

\section{Related Work} \label{section:related_works}
The fuzzy mining algorithm was introduced in \cite{gunther2007fuzzy}. The process model is built on the initial model with various 
filterings and abstractions. The initial model is the high-level description of processes that preserves all relations. We tried to use fuzzy and inductive mining traditionally for intrusion detection by performing conformance checking, but the result is far worse than expected \cite{van2005process}. We get the inspiration from fuzzy mining and modified the algorithm to perform online mining, which can then be used as a preprocessing step for online intrusion detection. The algorithm will be described in Section \ref{section:frequency}.

CICFlowMeter is a preprocessing tool that generates features, such as bytes per second, inter arrival time, packets per second etc., based on the network flows. The TCP flow terminates at FIN packet, where for UDP connections a timeout value needs to be set. The CICFlowMeter has been introduced in \cite{lashkari2017characterization}. We compare the performance of CICFlowMeter with our preprocessing algorithm in Section \ref{section:results}.

We use multi-layer perceptron (MLP), long short-term memory (LSTM), convolutional neural network (CNN), and k-nearest neighbours (KNN) in our binary classification and multi-class classification setups. The reason we choose these classifiers is:
\begin{itemize}
    \item MLP is a simple feedforward model.
    \item LSTM is a recurrent model that can be applied on time series data.
    \item CNN works directly on 2D inputs and it has been widely used in image classification problems.
    \item KNN is an example of traditional distance based classifier.
\end{itemize}

For the anomaly detection setup, we used the following outlier detectors.
\begin{itemize}
    \item Multivariate normal distribution (MND).
    \item Copula-Based Outlier Detection (COPOD).
    \item AutoEncoder.
    \item Angle-Based Outlier Detection (ABOD).
    \item Clustering-Based Local Outlier Factor (CBLOF).
    \item Histogram-based Outlier Score (HBOS).
    \item Isolation Forest (IForest).
    \item K-nearest Neighbors (KNN).
    \item Local Outlier Factor (LOF).
    \item Principal Component Analysis (PCA).
\end{itemize}

Note that the reason we choose these models as our classifiers is because they have different properties and characteristics, and the purpose of comparing them is just to verify that the preprocessing algorithms works as expected so it can be applied onto different classifiers.

\section{Dataset}

The dataset we used in this experiment is the CSE-CIC-IDS2018 dataset \cite{unb}. The dataset contains common attacks such as Bruteforce, DoS, and Botnet etc. The dataset comes with two formats, the extracted features in CSV spreadsheets and the PCAP binary packet data. We used Tshark to extract necessary attributes (IPs, ports, and flags) of TCP packets from the PCAP data for our algorithm. We also generated dataset with CICFlowMeter using the same PCAP data that were used as the training set for our preprocessing algorithm.

\section{Process Mining \& Measuring Frequency of Transitions} \label{section:frequency}
The packets observed on the wire is the sequence $P=\langle p_i \rangle_{i=1}^{n}$, where $p_i$ is each individual packet. The observed packets can also form a set of TCP flows $T=\{t_i\}_{i=1}^{m}$, where each flow $t_i$ can be constructed according to the IP addresses and ports of two hosts ($T$ can also be considered as the event log from a perspective of process mining). Please note that in process mining, a flow would correspond to 
a \emph{trace}, and both of these terms may be used in this paper interchangeably. We define that a new TCP flow is started when a packet that has flag SYN set but without ACK set (first packet of three-way handshake) is received, also, this initial packet determines the IP addresses and ports of two hosts. For example, the packet has $Source\ IP:Port=IP_1:PORT_1$ and $Target\ IP:Port=IP_2:PORT_2$. The bidirectional flows can be reconstructed based on forward direction ($IP_1:PORT_1\xrightarrow{}IP_2:PORT_2$) and backward direction ($IP_2:PORT_2\xrightarrow{}IP_1:PORT_1$). We define the TCP flow as completed when the packet that has the FIN flag or RST flag set is received.

As mentioned above, instead of analysing the flows, we analyse the relations between packets in flows, which is the basic idea of process mining \cite{van2011process}. Before we discuss the algorithm we need to define the concepts of \emph{transitions} and \emph{event classes}. 

Given a sequence of packets $P$, we define a \emph{transition} in $P$ as a pair of consecutive 
packets $\left(p_{i},p_{j}\right)$ within a flow in $P$. 

Here is an example, giving two traces $t_1$ and $t_2$, where $t_1 = \langle p_1,p_3,p_5\rangle $ and flow $t_2 = \langle p_2,p_6\rangle$, we will get two transitions for $t_1$: $\left(p_1,p_3\right)$ and $\left(p_3,p_5\right)$; one transition for $t_2$: $\left(p_2,p_6\right)$. Packets $p_1$ and $p_2$ come from two consecutive packets, however, these packets will not be considered as a transition as they belong to different flows.

An \emph{event class} $ec(p)$ of a packet $p$ is the concatenation of enabled flags of a packet followed by an indicator. e.g. 000.SYN.$|$C, where the last character is an indicator that indicates either the packet is sent from the client or the server. In this case, C indicates the packet is sent from the client.

A \emph{type} of the transition $(p_{i},p_{j})$ is a pair of corresponding event classes $(ec(p_{i}),ec(p_{j}))$. We will also refer to types of transitions as \emph{relations}. e.g. (000\-.SYN.$|$C, 000.ACK.SYN.$|$S) is a \emph{relation} and which indicates that a packet has SYN flag enabled is followed by a consecutive packet that has ACK and SYN enabled.

We have 23 possible event classes that were observed from the IDS2018 dataset of normal traffic data. We assume these 23 event classes cover the majority of possible flag combinations. All other packets that have flag combinations that were not observed in the dataset can be simply classified as OTHERS as a default rear case handling, or the event classes can be adjusted according to a particular situation.

There are 26 event classes in total, including 23 event classes from the flag data and 3 default classes (START, END and OTHERS). Therefore, there will be $26^2=676$ possible \emph{relations} if we assume every classes can be paired with other classes. These 26 event classes are available in Table \ref{tab:all_event_classes}. 

Our proposed online algorithm operates as follows. Given a sequence of packets $P$ (even log), the algorithm outputs the sequence (stream) of frequencies of relations observed in the last $l$ packets (for some $l$), organized in a form of adjacency matrix (26x26). Here, the frequencies of relations observed in the last $l$ packets are process models.

We want to measure the frequency by counting the incoming relations into an adjacency matrix $A$, however, we will only calculate the frequency based on the last $l$ packets, and the frequency of the transitions needs to be updated per each new packet.

If the initial packet $p_1$ in trace $t_1$ carries 000.SYN.$|$C, then the weight of $A(START, 000.SYN.|C)$ will be increased by one; and if the next packet $p_3$ in the same trace carries 000.ACK.SYN.$|$S, then the weight in $A(000.SYN.|C, \\000.ACK.SYN.|S)$ will be increased by one.

\begin{table}[h]
\caption{Possible event classes used.}
\resizebox{\textwidth}{!}{%
\centering
\begin{tabular}{|l|l|l|l|}
\hline
000.SYN.$|$C         & 000.ACK.SYN.$|$S     & 000.ACK.$|$C         & 000.ACK.PSH.$|$C     \\ \hline
000.ACK.PSH.$|$S     & 000.ACK.FIN.$|$C     & 000.ACK.$|$S         & 000.ACK.FIN.$|$S     \\ \hline
000.ACK.RST.$|$C     & 000.ACK.RST.$|$S     & 000.RST.$|$S         & 000.ACK.PSH.FIN.$|$S \\ \hline
000.RST.$|$C         & 000.CWR.ECE.SYN.$|$C & 000.ECE.ACK.SYN.$|$S & 000.NS.ACK.FIN.$|$S  \\ \hline
000.ACK.PSH.FIN.$|$C & 000.CWR.ACK.PSH.$|$C & 000.CWR.ACK.$|$C     & 000.CWR.ACK.$|$S     \\ \hline
000.CWR.ACK.PSH.$|$S & 000.CWR.ACK.RST.$|$S & 000.CWR.ACK.RST.$|$C & START                \\ \hline
END                  & OTHERS               &                      &                      \\ \hline
\end{tabular}
}
\label{tab:all_event_classes}
\end{table}

In our experiments, we have limited the number of packets $l$ that used to calculate the frequency of transitions to 500 by using the sliding window. The limitation here is just our choice based on various experiment and any number is possible to be used here. The sliding window starts from $p_1$ and covers $\langle p_i\rangle_{i=1}^{l}$, and the frequency of transitions $A'$ is calculated as $A/l$, then the window will be shifted one step further which covers $\langle p_i\rangle_{i=2}^{l+1}$. This process results in a sequence $\langle A'_i \rangle_{i=1}^{n-l+1}$ and each $A'_i$ is a snapshot of a process model with $l$ events. In process mining, the events are instances of event classes. The process of producing $A$ is similar to mining the fuzzy model where $t_i$ is traces and $P$ is the event log. However, the modification here is that the last state that is outside the window of each flow $t_i$ was kept in the state table so the START and END tokens will only occur at the beginning and end of a particular TCP flow, not where it begins and end in each sliding window. As the last state is known, the relation can be mined even if the window has already passed the previous event. In other words, the original process mining takes the entire event log $P$ into account, and in that case $P$ generates a single huge adjacency matrix $A$. However, this is not suitable for online processing, therefore, we keep the states of traces through the entire event log $P$ but limit the process model generation based on $l$ packets only. This keeps all the transition information and makes it suitable for online systems.

\begin{figure}[h]
	\centering
	\includegraphics[width=3.1in]{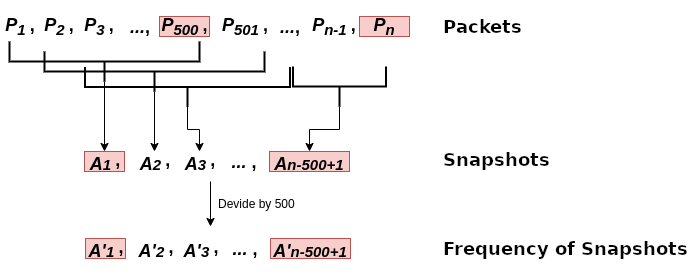}
	\caption{Diagram of algorithm \ref{algo:algo1}. Packets $P_{500}$ and $P_n$ belong to traces that are marked as attacks, therefore, $A_1$ and $A_{n-500+1}$ are also labelled as attacks for training classifiers.}
	\label{fig:diagram_algorithm}
\end{figure}

\begin{figure}[h]
	\centering
	\includegraphics[width=2.3in]{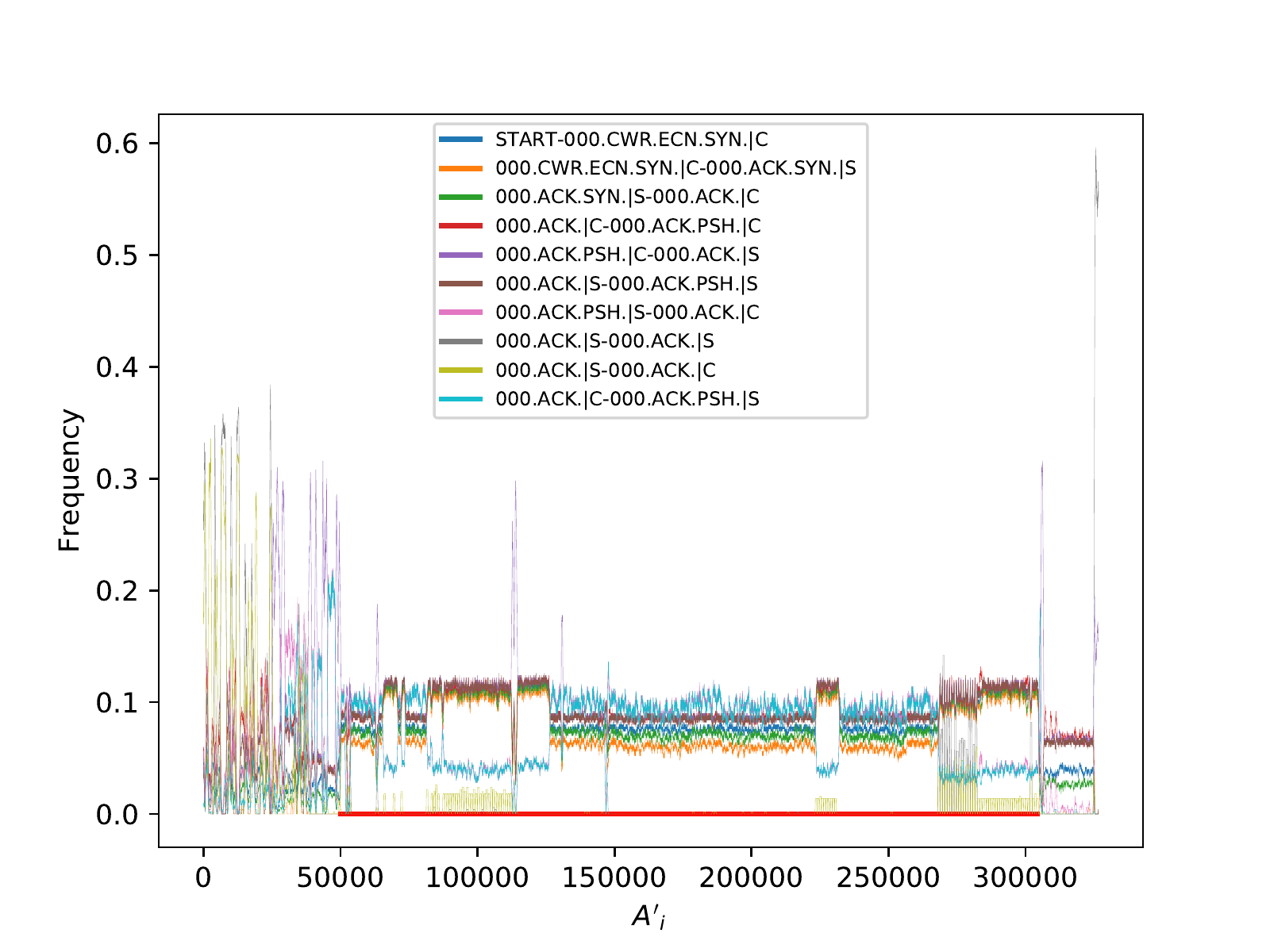}
	\includegraphics[width=2.3in]{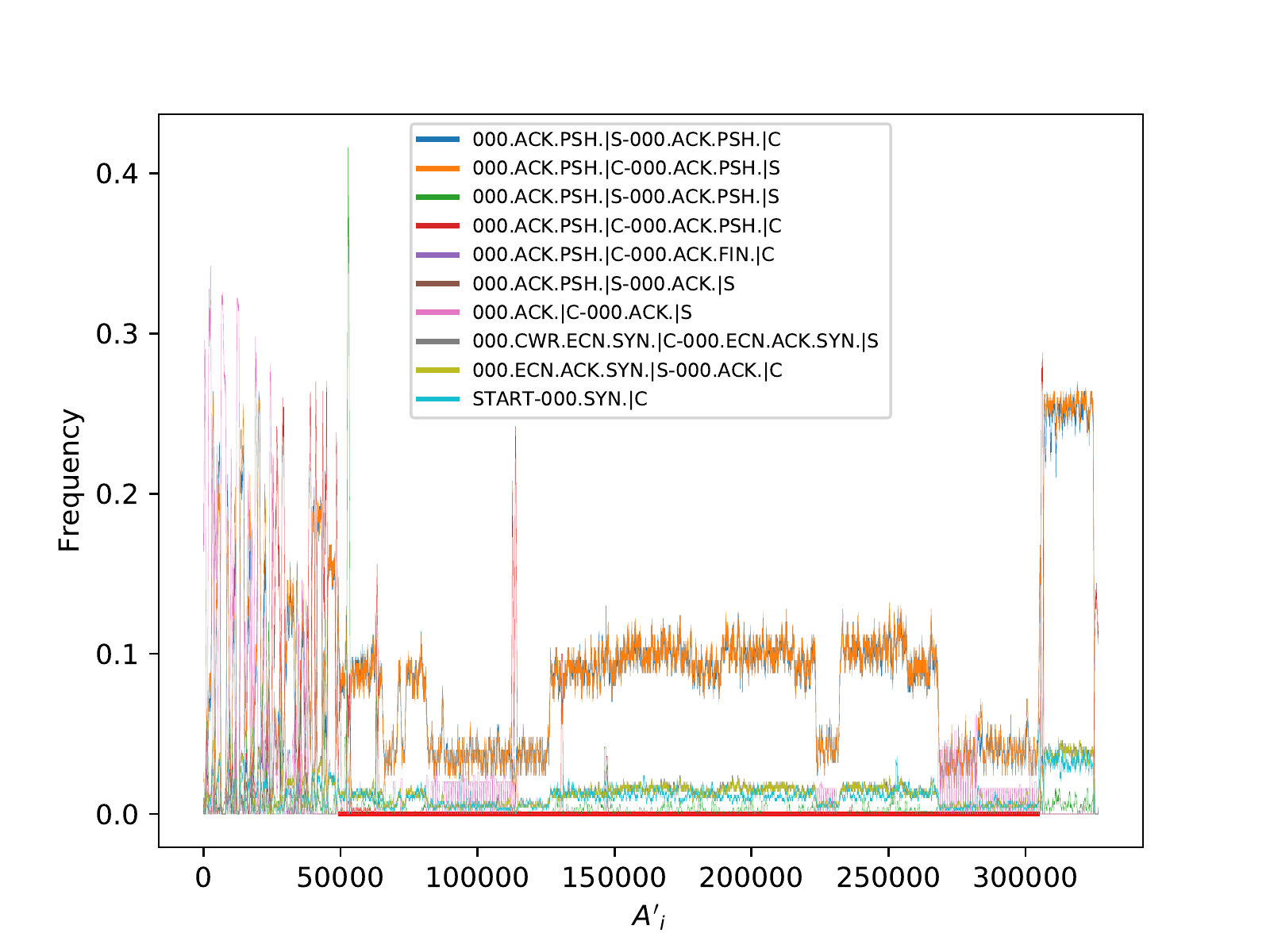}
	\caption{The chart show frequency fluctuation under Botnet attacks. 20 out of 676 possible relations are given as the example. 10 relations in the first chart and 10 relations in the second chart. In the first chart, the attack happens between $50,000^{th}$ to $310,000^{th}$ packets (ground truth), and it's clear that the frequency starts to stabilise.}
	\label{fig:sequence_of_a}
\end{figure}

For the purpose of performance and ease of use, we used the $l$-sized buffer to keep the transitions that are inside the sliding window instead of count every transition again in each loop, and we ignored the transition to END, therefore, we only need to update two elements in $A$ for each packet (decrease the frequency for the packet that goes off the buffer and increase the frequency for the packet that goes on the buffer). When computing the frequencies of transitions, we also produced the output for machine learning based on the labelled data provided from the dataset. The dataset provides the source of a certain attack, therefore, if the most recent packet in $A_i$ was sent from or to the IP that was labelled as a certain attack (13 types of attacks in total), $A_i$ will be marked as a model that contains attack. For better demonstration, Fig. \ref{fig:diagram_algorithm} of Botnet attacks is provided, where attacks are marked in red colour. The red bar at the bottom of the chart in Fig. \ref{fig:sequence_of_a} indicates the locations of attacks. In summary, we have two outputs from this process, the frequencies of the transitions $A'_i$ and the location of attacks. The pseudocode (Algorithm \ref{algo:algo1}) is given in Appendix \ref{pseudo} below and Fig. \ref{fig:sequence_of_a} shows the example of the output of 20 relations. 

\section{Experiments} \label{section:ml}
\subsection{Knowledge-based 
detection
}
The proposed workflow applicable in knowledge-based (signature-based) IDS is as follows. The IP and flag information of network packets are captured, and the flows are reconstructed according to the IPs. The flows go through our preprocessor, then a series of snapshots $A'_i$ matrices are generated. These snapshots will be reshaped if needed and then get fed into the ML classifiers trained to recognize \emph{known} attacks. Finally, the classifier will raise the alarm whenever we got an attack.

Except for the input of CNN, $A'_i$ ($26$ by $26$ matrices) was flattened into 676-dimensional vectors (the elements of the vector are the frequencies of transitions), and which was used as the input for MLP, LSTM, and KNN. For binary classifications, the output data are 2-D vectors where $\left[1 \ 0\right]^\intercal$ stands for normal traffic and $\left[0\ 1\right]^\intercal$ stands for intrusions; for multi-class classifications, the output was 14-dimensional one-hot vectors that encode the output to be normal or one of 13 types of attacks. As all of the locations of attacks were marked, the output data can be generated from the algorithm above (Section \ref{section:frequency}). We have not reduced the dimension as we want the classifiers to handle the input data.

The multi-layer perceptron was constructed with 4 dense layers (676, 128, 128 and 2 units for each layer) followed by a softmax layer; the LSTM was built with 2 layers of LSTM (128 units each), one dense layer with 2 units and one softmax layer lastly; the CNN model is similar to the model used in \cite{kim2019intrusion}, however, we increased the input shape to 26-by-26 and added one dense layer (32 units) before the output layer. For multi-class classification, the number of units for the last dense layer had been increased from 2 to 14 for all neural networks. All instances had been tested through 5-fold cross-validation. We have three setups for the LSTM with 50, 100 and 250 timesteps.

The dataset has imbalanced normal and anomalous entries. Therefore, normal data has been added into samples or been removed to match the number of the attack data for binary classifications. This helps to train the neural networks but does not affect the F-score. Also, attack types SQL-Injection and Infiltration have been discarded in binary classifications due to an insufficient amount of data.

\subsection{Anomaly-based detection}

The proposed workflow for anomaly-based IDS is as follows. The necessary information of network packets are captured, and flows are reconstructed, then the flows go through the preprocessor, which generates snapshots. These steps are identical to the signature-based IDS. The snapshots are reshaped into vectors and fed into outlier detectors, then the outlier detectors provide the outlier scores, where a higher outlier score indicates the data has a higher probability of being an anomaly intrusion.

We use the PyOD python library for anomaly-based detection, and hyperparameters of all outlier detectors remain default. The data format is the same as the one used in the binary classification, where all data have been reshaped to 676-dimensional vectors. For data generated with CICFlowMeter, we did the column-wise normalisation before feeding them into the detectors. We only use the normal data that do not contain any attack to train the detectors, then used mixed data, which contain both normal data and attack data for testing.

\section{Results} \label{section:results}
We compared the results of binary classification in Table \ref{tab:binary_classification_comparison}, and the results of binary classification have also been compared with the results from \cite{kim2019intrusion} (Table \ref{tab:binary_classification_CNN_comparison}). The LSTM gives worse result compared to other models, especially in multi-class classification (Table \ref{tab:multi_classification_comparison}). Some F1 scores produced by LSTM multi-class classification display NaN (not a number) or 0, meaning these classes have 0 in both precision and recall (i.e. 0 on the diagonal line in the confusion matrix), or 0 in either precision or recall.
However, this may indicates that with the preprocessing step, the classifiers do not require the historical data to perform the classification, as the historical data have already been encoded during the process showing in Fig. \ref{fig:diagram_algorithm}.

We focus on the preprocessing step, so we compare our results with CICFlowMeter in Table \ref{tab:binary_classification_CNN_comparison}. Both column uses CNN as the classifier, and the result of our approach is promising.

\begin{table*}[h]
    \caption{the F1 scores for binary classification.}
    \centering
    \begin{tabular}{|l|l|l|l|l|l|l|l|}
    \hline
    \textbf{Attack} & \textbf{MLP} & \textbf{LSTM-50} & \textbf{LSTM-100} & \textbf{LSTM-250} & \textbf{KNN} & \textbf{CNN} \\ \hline
    FTP-BruteForce & 0.9990          & 0.9976          & 0.9984          & 0.9982   & \textbf{0.9991} & 0.9990  \\ \hline 
    SSH-Bruteforce & 0.9763          & 0.9764          & 0.9763          & 0.8907   & 0.9732          & \textbf{0.9764}  \\ \hline 
    DoS-GoldenEye  & 0.9212          & 0.9680          & 0.7457          & 0.9498   & \textbf{0.9821} & 0.9434 \\ \hline 
    DoS-Slowloris  & 0.9945          & 0.8490          & 0.8513          & 0.7770   & 0.9947          & \textbf{0.9948} \\ \hline 
    DoS-SlowHTTP   & 0.9983          & 0.9798          & 0.9937          & 0.9803   & \textbf{0.9985} & 0.9984 \\ \hline 
    DoS-Hulk       & 0.7309          & 0.6732          & \textbf{0.7731} & 0.7313   & 0.7381          & 0.7314 \\ \hline 
    DDoS-LOIC-HTTP & \textbf{0.9968} & 0.7606          & 0.7129          & 0.7443   & 0.8353          & 0.8406 \\ \hline 
    DDOS-HOIC      & \textbf{0.9687} & 0.7072          & 0.7922          & 0.6935   & 0.7879          & 0.7559 \\ \hline 
    BruteForce-Web & \textbf{0.9962} & 0.9588          & 0.9621          & 0.9631   & 0.9789          & 0.9741 \\ \hline 
    BruteForce-XSS & \textbf{0.9985} & 0.9676          & 0.9674          & 0.9788   & 0.9868          & 0.9827 \\ \hline 
    Botnet         & 0.8168          & \textbf{0.9644} & 0.9137          & 0.8308   & 0.8754          & 0.8623 \\ \hline 
    \end{tabular}%
    \label{tab:binary_classification_comparison}
\end{table*}

\begin{table*}[h]
    \caption{the F1 scores for multi-class classifications.}
    \centering
    \begin{tabular}{|l|l|l|l|l|l|l|}
    \hline
    \textbf{Attack Type} & \textbf{MLP} & \textbf{LSTM-50} & \textbf{LSTM-100} & \textbf{LSTM-250} & \textbf{KNN} & \textbf{CNN} \\ \hline
    Normal         & 0.5533          & 0.4117  & 0.5401   & 0.3115   & \textbf{0.8306} & 0.6457          \\ \hline
    FTP-BruteForce & 0.9734          & NaN     & NaN      & NaN      & \textbf{0.9991} & 0.9743          \\ \hline
    SSH-Bruteforce & 0.9772          & 0.9254  & 0.9206   & 0.9215   & 0.9753          & \textbf{0.9774} \\ \hline
    DoS-GoldenEye  & 0.9287          & 0.9219  & 0.9209   & 0.9215   & \textbf{0.9618} & 0.9277          \\ \hline
    DoS-Slowloris  & 0.9830          & 0.2433  & 0.9573   & 0.8323   & \textbf{0.9935} & 0.9837          \\ \hline
    DoS-SlowHTTP   & 0.9070          & 0.5056  & 0.5042   & 0.5546   & \textbf{0.9981} & 0.8844          \\ \hline
    DoS-Hulk       & 0.8241          & 0.8245  & 0.8212   & 0.7611   & 0.7688          & \textbf{0.8244} \\ \hline
    DDoS-LOIC-HTTP & \textbf{0.8755} & 0.8754  & 0.524    & 0.8668   & 0.8480          & \textbf{0.8755} \\ \hline
    DDOS-HOIC      & 0.8188          & 0.8232  & 0.2553   & 0.8078   & \textbf{0.8253} & 0.8203          \\ \hline
    BruteForce-Web & 0.9659          & 0.0334  & NaN      & 0        & 0.9679          & \textbf{0.9687} \\ \hline
    BruteForce-XSS & 0.9634          & NaN     & NaN      & 0        & \textbf{0.9847} & 0.9756          \\ \hline
    SQL-Injection  & 0.1765          & NaN     & NaN      & NaN      & \textbf{0.4941} & 0.2908          \\ \hline
    Infiltration   & 0.0898          & 0.0417  & 0.142    & 0.0179   & \textbf{0.4334} & 0.0149          \\ \hline
    Botnet         & 0.6473          & 0.6465  & 0.6295   & 0.4089   & 0.6277          & \textbf{0.6540} \\ \hline
    \end{tabular}%
    \label{tab:multi_classification_comparison}
\end{table*}

\begin{table}[h]
    \caption{comparison between preprocessors.}
    \centering
    \begin{tabular}{|l|l|l|}
    \hline
    \textbf{Attack Type} & \textbf{Our Preprocessor} & \textbf{CICFlowMeter} \\ \hline
    FTP-BruteForce & \textbf{0.9990}  & 0.98         \\ \hline
    SSH-Bruteforce & \textbf{0.9764}  & 0.96         \\ \hline
    DoS-GoldenEye  & \textbf{0.9434}  & 0.47         \\ \hline
    DoS-Slowloris  & \textbf{0.9948}  & 0.66         \\ \hline
    DoS-SlowHTTP   & 0.9984           & \textbf{1}   \\ \hline
    DoS-Hulk       & 0.7314           & \textbf{1}   \\ \hline
    DDoS-LOIC-HTTP & 0.8406           & \textbf{1}   \\ \hline
    DDOS-HOIC      & 0.7559           & \textbf{1}   \\ \hline
    BruteForce-Web & \textbf{0.9741}  & 0.3          \\ \hline
    BruteForce-XSS & \textbf{0.9827}  & 0.65         \\ \hline
    Botnet         & 0.8623           & \textbf{1}   \\ \hline
    \end{tabular}
    \label{tab:binary_classification_CNN_comparison}
\end{table}

\begin{figure}[h]
	\centering
	\includegraphics[width=2.3in]{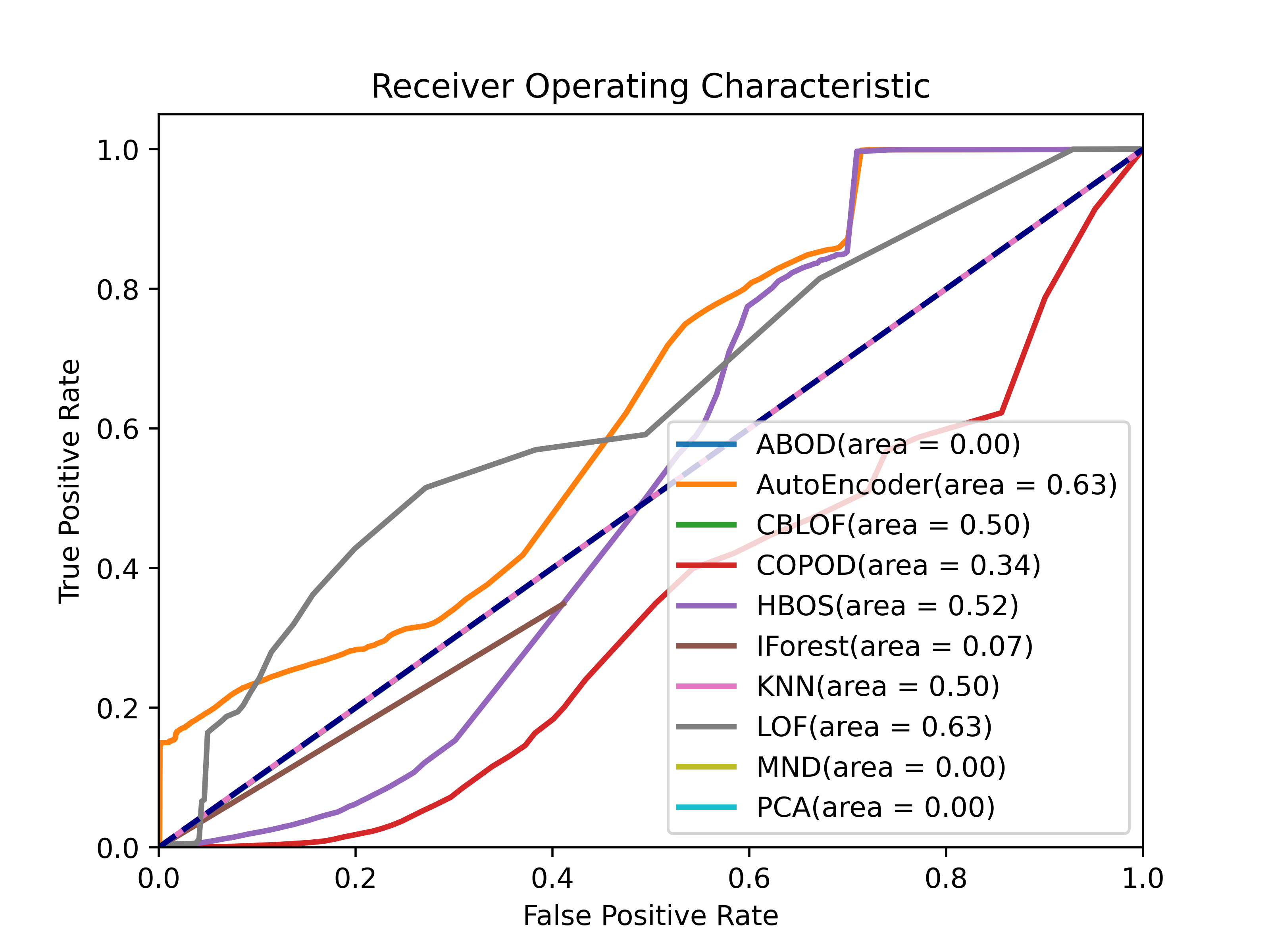}
	\includegraphics[width=2.3in]{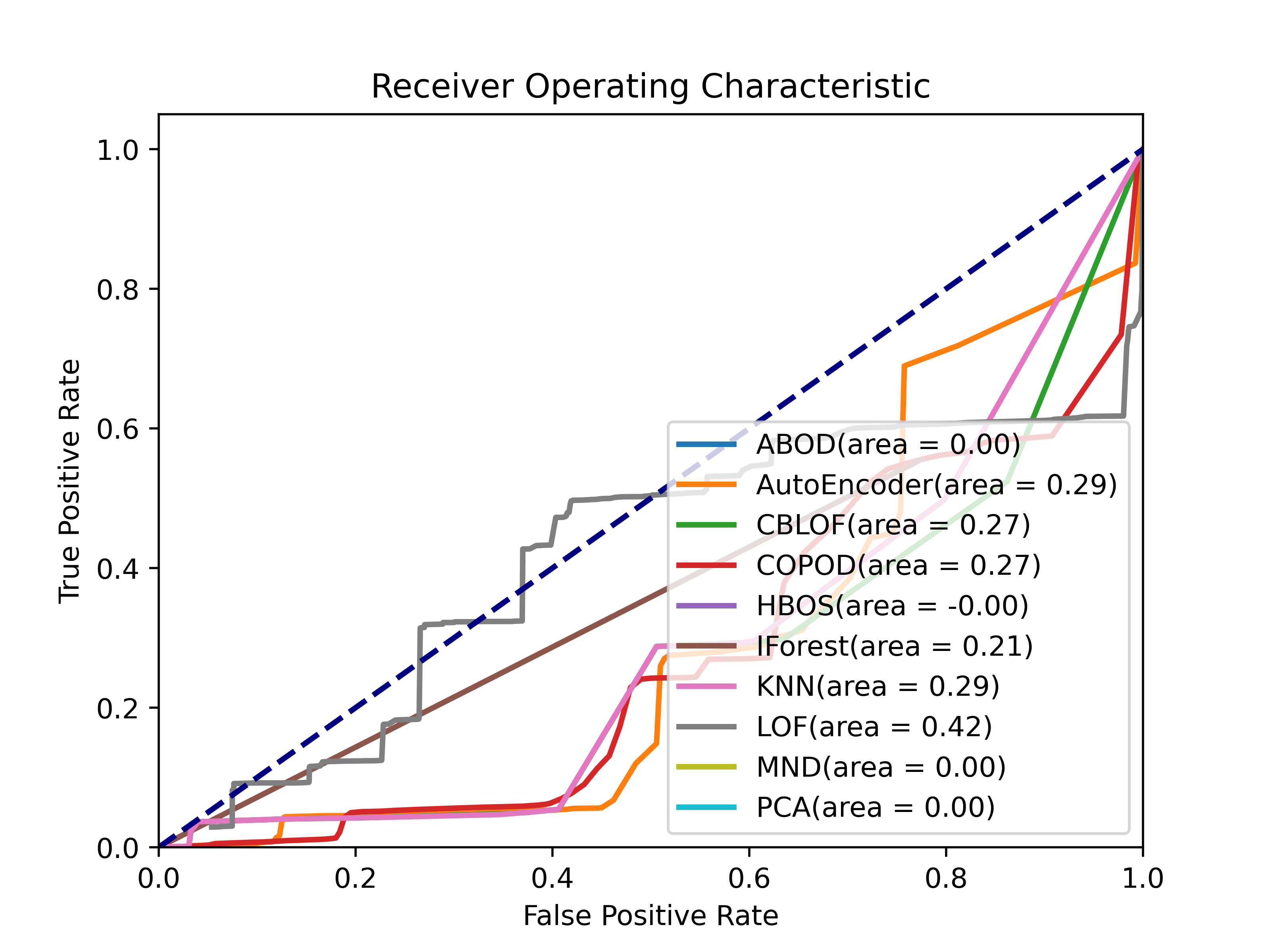}
	\caption{The receiver operating characteristic (ROC) for anomaly-based intrusion detection setup. The first chart shows the performance of our preprocessor, where the second chart shows the performance of CICFlowMeter.}
	\label{fig:roc}
\end{figure}

We compared the receiver operating characteristic (ROC) for the anomaly-based intrusion detection setup in Fig. \ref{fig:roc}. The type of attacks was not separated, so what we have here is the overall performance. It's clear that our algorithm works better in anomaly-based intrusion detection.

We have published the preprocessed data and some of the experiment results online.\footnote{https://zenodo.org/record/5616678}

\section{Conclusion \& Discussion} \label{section:conclusion}
We proposed the online process mining algorithm that preprocesses packet data for intrusion detection. The initial process mining algorithm was modified to adapt online process mining, which produces a series of process model snapshots with fixed window sizes. The snapshots were normalized and being used as the input data for machine learning. For signature-based intrusion detection, we used several machine learning models for binary and multi-class classification and yielded high accuracy. Though our preprocessing algorithm does not produce better results for every aspect, the performance was consistently high. On the anomaly-based intrusion detection side, the result is not impressive; however, considering the nature of anomaly-based intrusion detection and the performance of CICFlowMeter, we are satisfied with our preprocessing algorithm. We take it as a starting point to further extend the research on process mining applications in intrusion detection. 

Currently, we use snapshots similar to the initial model of the fuzzy miner, and it is possible to apply abstraction on the snapshots easily; therefore, our model can be easily extended with process mining techniques.

As mentioned in Section \ref{section:results}, the classification step does not require any historical snapshots as the historical information has already been packed into the latest snapshot. This might be more efficient than using a recurrent neural network (RNN) to directly classify the latest $n$ packet data without separate the flows; and it might be more accurate as our approach keeps the state of the flow until connection closed, not just getting information from the last $n$ packets. It needs to be clarified here that this is just a hypothesis.

Because this algorithm is used for preprocessing, other methods for classification can be easily applied. The output of our preprocessor is normalised and can be directly fed into many classifiers or outlier detectors without much modification. Furthermore, the only attribute we used to generate event classes is the flag, so there is still a large set of unconsidered attributes we could use. These could be verified in our future research. Another problem for machine learning on IDS is that when training the neural networks with one dataset but test the accuracy with a different dataset, the accuracy drops massively \cite{al2018re}. We are planning to verify if process-mining based preprocessing can help to resolve such an issue.
%
\appendix
    \section{Pseudocode} \label{pseudo}

    \begin{algorithm}[htbp]
    \SetKw{KwPush}{push}
    \SetKw{KwPop}{pop}
    \SetKw{KwInto}{into}
    \SetKw{KwIn}{in}
    \SetKw{KwOr}{or}
    \SetKw{KwAnd}{and}
    \SetKw{KwNot}{not}
    \SetKw{KwAppend}{append}
    \scriptsize
    \SetAlgoLined
    	\KwIn{$P=[n]$\tcc*[r]{load $n$ packets}}
    	\KwIn{$l=500$\tcc*[r]{define the window size}}
    	{$A=\left[26\;by\;26\right]$\tcc*[r]{a $26*26$ adjacency matrix}}
    	{$list\_A'=[\;]$\tcc*[r]{initialise list of $A'_i$}}
    	{$list\_attacks=[\;]$\tcc*[r]{initialise list of attacks}}
    	
    	\tcc{a dictionary where the key is the concatenation of IPs and Ports ($''IP_1:PORT_1|IP_1:PORT_1''$) of hosts, and the value is the flags of the previous packet}
    	{$dict\_state\_table=\{ \}$\;}
    	
    	{$buffer=[l]$\tcc*[r]{an FIFO buffer that keeps the last $l$ transitions (events)}}
    	\tcc{initialise with first $l$ packets}
     	\For{$i=1$ \KwTo $l$}
     	{
     		\tcc{check if the packet belong to any existing flow}
      		\uIf{$''IP_1:PORT_1|IP_2:PORT_2''$ \KwIn $dict\_state\_table.key()$ \KwOr $''IP_2:PORT_2|IP_1:PORT_1''$ \KwIn $dict\_state\_table.key()$}
      		{
      			\tcc{count the transition into $A$}
      			$A[dict\_state\_table[''IP_1:PORT_1|IP_2:PORT_2''],\ current\_flags]\mathrel{+}=1$\;
      			\tcc{update the state of the flow to the current flags into the dict}
       			$dict\_state\_table[''IP_1:PORT_1|IP_2:PORT_2''] = current\_flags$\;
       			\KwPush $current\_flags$ \KwInto $buffer$\;
       		
       			\tcc{check whether TCP flow terminates}
       			\If{$''FIN''$ \KwIn $current\_flags$ \KwOr $''RST''$ \KwIn $current\_flags$}
       			{
       				remove key $''IP_1:PORT_1|IP_2:PORT_2''$ from $dict\_state\_table$\;
       			}
       		}
       		\tcc{check if new TCP flow starts}
       		\ElseIf{$''SYN''$ \KwIn $current\_flags$ \KwAnd $''ACK''$ \KwNot \KwIn $current\_flags$}
       		{
       			$A[dict\_state\_table[''START''],\ current\_flags]\mathrel{+}=1$\;
       			\KwPush $current\_flags$ \KwInto $buffer$\tcc*[r]{push the transition (event) into buffer}
       		}
    	}
    	
    \caption{pseudocode of packet preprocessing.}
    \label{algo:algo1}
    \end{algorithm}
    
    \begin{algorithm}[htbp]
    \SetKw{KwPush}{push}
    \SetKw{KwPop}{pop}
    \SetKw{KwInto}{into}
    \SetKw{KwIn}{in}
    \SetKw{KwOr}{or}
    \SetKw{KwAnd}{and}
    \SetKw{KwNot}{not}
    \SetKw{KwAppend}{append}
    \setcounter{AlgoLine}{20}
    \scriptsize
    \SetAlgoLined
    
    	\KwAppend $A/l$ \KwTo $list\_A'$\tcc*[r]{append the frequency of transitions into the list}
    	
    	\For{$i=l+1$ \KwTo $n$}
     	{
      		\uIf{$''IP_1:PORT_1|IP_2:PORT_2''$ \KwIn $dict\_state\_table.key()$ \KwOr $''IP_2:PORT_2|IP_1:PORT_1''$ \KwIn $dict\_state\_table.key()$}
      		{
      			$A[\KwPop \ buffer]\mathrel{-}=1$\tcc*[r]{sub 1 for transition that went outside the window}
      			$A[dict\_state\_table[''IP_1:PORT_1|IP_2:PORT_2''],\ current\_flags]\mathrel{+}=1$\;
       			$dict\_state\_table[''IP_1:PORT_1|IP_2:PORT_2''] = current\_flags$\;
       			\KwPush $current\_flags$ \KwInto $buffer$\;
       		
       			\tcc{check whether TCP flow terminates}
       			\If{$''FIN''$ \KwIn $current\_flags$ \KwOr $''RST''$ \KwIn $current\_flags$}
       			{
       				remove key $''IP_1:PORT_1|IP_2:PORT_2''$ from $dict\_state\_table$\;
       			}
       		}
       		\tcc{check if new TCP flow starts}
       		\ElseIf{$''SYN''$ \KwIn $current\_flags$ \KwAnd $''ACK''$ \KwNot \KwIn $current\_flags$}
       		{
       			$A[\KwPop \ buffer]\mathrel{-}=1$\;
       			$A[dict\_state\_table[''START''],\ current\_flags]\mathrel{+}=1$\;
       			\KwPush $current\_flags$ \KwInto $buffer$\;
       		}
       		
    		\KwAppend $A/l$ \KwTo $list\_A'$\;
    		\tcc{Attack IP is from the labelled data}
    		\If{Attack IP \KwIn $current\_flags$}
    		{
    			\KwAppend $i$ to $list\_attacks$
    		}
    	}
    	
    	\KwOut{$list\_A'$\;}
    	\KwOut{$list\_attacks$\;}
    
    \end{algorithm}

\clearpage

\bibliographystyle{splncs04}
\bibliography{main}

\end{document}